\documentclass[journal]{IEEEtran}
\usepackage{latexsym}
\usepackage{amsfonts}
\usepackage{amssymb}
\usepackage{amsbsy}
\usepackage{amsmath}
\usepackage{amsthm}
\usepackage[belowskip=-10pt,aboveskip=0pt]{caption}
\usepackage{graphicx,subfig}
\usepackage[varg]{txfonts}
\usepackage[hidelinks,implicit=false]{hyperref}
\usepackage{algorithm,algpseudocode}

\title{\LARGE An Iterative Algorithm for Optimal Carrier Sensing Threshold\\
in Random CSMA/CA Wireless Networks}
\author{Dong Min Kim and Seong-Lyun Kim
\IEEEcompsocitemizethanks{\IEEEcompsocthanksitem The authors are
with the Radio Resource Management and Optimization Laboratory,
School of Electrical and Electronic Engineering, Yonsei University,
50 Yonsei-ro, Seodaemun-gu, Seoul 120-749, Korea (email:
dmkim@ramo.yonsei.ac.kr; slkim@ramo.yonsei.ac.kr).}%
\thanks{This research was funded by the MSIP (Ministry of Science, ICT \& Future Planning), Korea in the ICT R\&D Program 2013
(1291101107-130010300).}
}

\begin{document}
\maketitle

\begin{abstract}
We investigate the optimal carrier sensing threshold
in random CSMA/CA networks considering
the effect of binary exponential backoff.
We propose an iterative algorithm for optimizing the carrier sensing threshold and hence maximizing
the area spectral efficiency.
We verify that simulations are consistent with our analytical results.
\end{abstract}


\vspace{-5mm}

\section{Introduction}

To enhance wireless connectivity and capacity,
efficient multiple access schemes for spatially randomly distributed nodes
are necessary.
The most widely used multiple access
scheme is Carrier Sense Multiple Access with Collision Avoidance (CSMA/CA).
In this letter, we propose an iterative algorithm for finding
the optimal carrier sensing threshold of spatially randomly distributed
CSMA/CA wireless networks.

The fundamental processes in CSMA/CA are carrier sensing and random backoff.
Carrier sensing provides the spatial resolution to concurrent transmitters
and random backoff gives the temporal resolution to concurrent transmitters
at the nearby place.
The IEEE~802.11 Distributed Coordination Function (DCF) \cite{802.11spec2007}
utilizes physical carrier sensing (optionally virtual carrier sensing) and
binary exponential backoff (BEB).
In physical carrier sensing,
such as the energy detection method, the node
senses the medium to measure the aggregate interference, and transmission can
begin only if the measured interference is below the carrier sensing threshold.
In virtual carrier sensing, the node that
intends to transmit performs proactive actions to prevent nodes in the vicinity
from transmitting simultaneously with it.
When the medium becomes idle, multiple transmitters would access simultaneously,
causing collisions.
By utilizing BEB, contention conflict is avoidable.

The carrier sensing threshold is a significant parameter to balance
the tradeoff between the spatial reuse and the packet collision
by controlling the aggregate interference.
In \cite{Yang2005},
it is noted that optimizing carrier sensing is
important to increase the throughput performance.
The authors of \cite{Yang2005}
investigate the optimal carrier sensing range under the
regular hexagonal topology, whereas
we consider the spatially randomly distributed interferers
to realistically capture the effect of interference
using stochastic geometry \cite{Stoyan1995}.

Some researches conducted to
determine the spatial distribution of transmitting nodes in the CSMA/CA network.
In one such work \cite{nguyen2007}, the authors applied the
Mat\'{e}rn hard-core process (MHP) \cite{Stoyan1995} to model the spatial
transmitter pattern. MHP is a dependent
thinning process of the Poisson point process (PPP) used to create
separation of the marked points by at least a certain minimum distance.
In \cite{Busson2009}, the authors proposed a simple
sequential inhibition (SSI) point process to model
for the same purpose, which is not mathematically tractable.
In \cite{nguyen2007} and \cite{Busson2009}, backoff scheme was not considered
and the optimal carrier sensing threshold is not provided.

Later, the authors of \cite{Alfano2011} investigated the
throughput performance of dense CSMA networks
from the stochastic geometry point of view.
However, the authors of \cite{Alfano2011} did
not find the optimal carrier sensing threshold and
did not consider the effect of random backoff either.
In \cite{Kaynia2011}, the authors investigated
the optimal carrier sensing threshold based on a
lower bound for the outage probability.
They considered the effect of one strong interference and simplified the backoff scheme,
whereas we consider the effect of aggregate interference and the effect of BEB,
such as, collisions in the contention period, increasing of the backoff interval and
backoff freezing behavior.

\vspace{-4mm}

\section{Problem Definition}

The \emph{area spectral efficiency} $\eta$ (ASE), which is defined as
the product of successfully transmitting node density and the data rate,
provides a framework to quantify the capacity of the wireless network \cite{Weber2010}.
Our problem is to find the optimal carrier sensing threshold $I_s^*$ that
maximizes ASE $\eta$ as follows:
\begin{equation}\label{E:ase}
I_s^* = \mathop {\arg \max }\limits_{{I_s}} {\text{ }}{\lambda _t}\log_2 \left( {1 + \beta } \right){p_s},
\end{equation}
where $\lambda_t$ denotes the active transmitter density in the contention free
period of CSMA/CA and $\beta$ means the target SIR.
The transmission success probability is denoted by $p_s$.

We propose an iterative algorithm for finding $I_s^*$ as described in Algorithm \ref{A:iter}.
We will explain how the proposed algorithm is obtained and show the
performance of the algorithm.

\begin{algorithm}
\caption{Proposed Algorithm.
}\label{A:iter}
\begin{algorithmic}[1]
\State Initialize $I_s^{\text{next}}$ with a small value less than $r_t^{-\alpha}P$
\State $I_s^{\text{current}} \gets I_s^{\text{next}} + 1$
\While{$I_s^{\text{next}} \not= I_s^{\text{current}}$}
    \State $I_s^{\text{current}} \gets I_s^\text{next}$
    \State $\tau_{\text{current}} \gets 1$, $\tau_{\text{next}} \gets 0$
    \While{$\tau_{\text{next}} \not= \tau_{\text{current}}$}
        \State $\tau_\text{current} \gets \tau_\text{next}$
        \State ${\tau_\text{next}} \gets {{\tau_\text{current}} - {\frac{(\tau_\text{current} - h(\tau_\text{current}))}{(1-h'(\tau_\text{current}))}}}$
        \Comment{$h$ is RHS of \eqref{E:tau}}
    \EndWhile\label{innerendwhile}
    \State $\tau \gets \tau_\text{next}$
    \State update $\eta$ with $\tau$
    \Comment{$\eta$ is \eqref{E:ase1}}
    \State ${I_s^{\text{next}}} \gets {I_s^{\text{current}}} - \frac{{\eta'} \left(I_s^{\text{current}}\right)}{{\eta''} \left(I_s^{\text{current}}\right)}$
\EndWhile\label{outerendwhile}
\State $I_s^* \gets I_s^{\text{next}}$ \Comment{get the optimal carrier sensing threshold}
\end{algorithmic}
\end{algorithm}

\vspace{-4mm}

\section{System Model}

\subsubsection{Topology and Channel Modeling}

Consider a wireless network, in which all transmitters
communicate with their receivers over a common wireless channel.
Transmitters are located according to a homogeneous PPP with intensity $\lambda$.
This kind of network topology is called the \emph{Poisson
bipolar network} \cite{Baccelli2009}.
Each transmitter $i$ has infinite backlogged data to transmit.
The transmitter/receiver
pairs vary over time, but we focus on a snapshot of the
overall communication process.
The channel gain from transmitter $i$ to receiver $j$ is modeled by
$g_{i,j}d_{i,j}^{-\alpha}$, where $g_{i,j}$ is
an independently and identically distributed
exponential random variable with unit mean,
which reflects the effect of Rayleigh fading.
The distance between nodes $i$ and $j$ is denoted by $d_{i,j}$ with
the path loss exponent $\alpha$. Using a common channel,
different communication pairs can interfere with one another.
Let $P$ be the transmit power and
an associated receiver $j$ is at a distance of $r_t$ from the
typical transmitter $i$.
Assuming the network is interference-limited
and the receiver noise is ignored,
then the signal-to-interference-ratio (SIR) $\gamma_{j}$
is given by:
\begin{equation}\label{E:sir}
\gamma_{j} = \frac{{g_{i,j}r_t^{ - \alpha }P}}
{{\sum\limits_{u \in {{{\cal T}}_i},j \notin {{{\cal T}}_i}} {g_{u,j}d_{u,j}^{ - \alpha } P }
}}={\frac{{{g_{i,j}r_t^{ - \alpha }P}}}{I}},
\end{equation}
where $I$ denotes the aggregate interference and ${{{\cal T}}_i}$
denotes the set of concurrently transmitting (interfering) nodes
when node $i$ transmits.
For a given target SIR $\beta$, a transmission succeeds
if $\gamma_{j}$ is greater than or equal to $\beta$.
The data rate of the typical transmitter $i$ is a function
of $\beta$.
We use Shannon's formula
$\log_2\left(1\!+\!\beta\right)$ in which we assume a unit bandwidth.

\subsubsection{CSMA/CA Modeling}

Let us assume that the network employs the CSMA/CA random access scheme,
especially, RTS/CTS mode \cite{802.11spec2007}.
In CSMA/CA with BEB,
if the channel is idle during the predetermined time
(DIFS in IEEE~802.11 DCF), the transmitters enter the contention period.
Each transmitter should defer its transmission
during a randomly selected slotted contention window.
The backoff counter is decremented in each slot time
if the channel is still sensed idle.
When the backoff counter is expired,
every contending communication pair exchanges
control packets (RTS/CTS) to reserve a wireless channel.
The transmitters who conducted this process successfully
enter the contention-free period, and proceed data transmissions.
If the transmission is failed,
the contention window size increases exponentially.

The seminal works of \cite{Bianchi2000} and \cite{Cali2000} show that
the effect of BEB can be appropriately modeled by
$p$-persistence medium access analysis.
We denote $\tau$ as a steady state medium access probability.
Each transmitter accesses the medium by Bernoulli trial with probability $\tau$.
Therefore, after the medium becomes idle, contending node density is
$\lambda\tau$. Let $p_c$ be the collision probability of control messages
in the contention period, and using the result of the \cite{Baccelli2009},
we obtain $p_c$ as follows:
\begin{equation}\label{E:pc}
{p_c} = 1 - \exp \left( { - \lambda \tau r_t^2\beta _c^{\frac{2}{\alpha }}
\frac{{2{\pi ^2}}}{{\alpha \sin \left( {2\pi /\alpha } \right)}}} \right),
\end{equation}
where $\beta_c$ denotes the target SIR for the control messages.

Let $P_b$ be the channel busy probability, which is the probability
that the aggregate interference is greater than or equal to a given
carrier sensing threshold $I_s$. In \cite{Souryal2001}, the authors derived
the cumulative distribution function of the interference
in PPP networks with Rayleigh fading. We modified their result with
a transmit power term $P$, then $p_b$ is
\begin{equation}\label{E:pb}
  {p_b} = \Pr \left[ {I \geq {I_s}} \right]
   = \operatorname{erf} \left( {\frac{{{\pi ^2}\lambda \tau }}{4}\sqrt {\frac{P}{{ {I_s}}}} } \right).
\end{equation}

In \cite{Ziouva2002csma},
the authors derived $\tau$
considering the effect of BEB,
such as, collision from overlapping of backoff counter,
increase of backoff window size and freezing of backoff counter.
However, they assumed a node can access the medium without backoff
after successful transmission.
This assumption is not compatible with the
IEEE 802.11 \cite{802.11spec2007}.
We newly derived $\tau$ by correcting the erroneous assumption
in \cite{Ziouva2002csma} as follows:\footnote{Due to space limitation,
we omit the derivation of \eqref{E:tau}. Please see
\url{http://hertz.yonsei.ac.kr/tau.pdf}.}
\begin{equation}\label{E:tau}
\tau  = \frac{{2\left( {1 - {p_b}} \right)\left( {1 - 2{p_c}} \right)}}{{\left( {1 - 2{p_c}} \right)\left( {1 - 2{p_b} + {W_0}{{\left( {2{p_c}} \right)}^m}} \right) + {W_0}\left( {1 - {p_c}} \right)\left( {1 - {{\left( {2{p_c}} \right)}^m}} \right)}},
\end{equation}
where $m$ is the maximum backoff stage
and $W_0$ is the initial backoff window size.
The value $\tau$ is a function of collision probability $p_c$ and
channel busy probability $p_b$.
By substituting \eqref{E:pc} and \eqref{E:pb} into \eqref{E:tau},
let the right-hand side of \eqref{E:tau} be $h\left(\tau\right)$.
We obtain the fixed point formulation $\tau=h\left(\tau\right)$, which
can be numerically solved by Newton's method as follows:
\begin{equation}\label{E:newton_tau}
\tau _{n + 1}  = \tau _n  - \frac{{\tau _n  - h\left( {\tau _n } \right)}}
{{1 - h^\prime  \left( {\tau _n } \right)}},
\end{equation}
where $\tau_n$ denotes the value of $\tau$ at $n$-th iteration and
$h'$ denotes a derivative of $h$ with respect to $\tau$.
The iterative method \eqref{E:newton_tau} works well with an initial value $\tau_0\!\!=\!\!0$.
The value $\tau$ is validated by simulations performed in NS-3
with various $\lambda$, $I_s$ and $\beta_c$.
In simulations, $W_0$ is 32 and increases up to 1024.
The transmitters continuously generate 1KB packets to model
the saturated traffic. The transmit power is 30dBm.
The simulation area is $10\text{km}\!\times\!10\text{km}$.
To model the PPP network,
the number of transmitters $N$ in the network
is generated according to the Poisson distribution with $\lambda\!\times\!10\text{km}\!\times\!10\text{km}$.
For example, if $\lambda$ is 0.0001nodes/$\text{m}^2$, the average number of nodes in the area is $10^4$,
where the transmitters are uniformly distributed.\footnote{A homogeneous Poisson point process
with $\lambda$ in infinite area
becomes a uniform distribution of $k$ nodes on the finite area of size $\mathcal{A}$,
where $k = \lambda\mathcal{A}$.}
The receivers
are located at a distance $r_t$ with random directions from transmitters.
The Rayleigh fading channel is modeled by the Nakagami propagation loss component
in NS-3 with proper parameter settings.
The simulation results are averaged over hundreds of simulation runs.
As shown in Table~\ref{T:tau}, the simulation results
are consistent with the analytical results.

\begin{table}[tb]
\centering
\caption{The medium access probability $\tau$.} \label{T:tau}
\begin{tabular}{|l@{}|@{}l@{}|@{}l@{}|@{}l@{}|@{}l@{}|@{}l@{}|@{}l@{}|@{}l@{}|@{}l@{}|@{}l@{}|@{}l@{}|@{}l@{}|@{}l@{}|}
\hline
$\lambda$ & \multicolumn{4}{c|}{0.0001} & \multicolumn{4}{c|}{0.001} & \multicolumn{4}{c|}{0.01} \\
\hline
$I_s$ (dBm) & \multicolumn{2}{c|}{-40} & \multicolumn{2}{c|}{-10} & \multicolumn{2}{c|}{-40} & \multicolumn{2}{c|}{-10} & \multicolumn{2}{c|}{-40} & \multicolumn{2}{c|}{-10} \\
\hline
$\beta_c$ (dB) & {\rm{\ }}3 & {\rm{\ }}10 & {\rm{\ }}3 & {\rm{\ }}10 & {\rm{\ }}3 & {\rm{\ }}10 & {\rm{\ }}3 & {\rm{\ }}10 & {\rm{\ }}3 & {\rm{\ }}10 & {\rm{\ }}3 & {\rm{\ }}10 \\
\hline
$\tau$ (simul.) &{\rm{\ }}.053  &{\rm{\ }}.043  &{\rm{\ }}.055  &{\rm{\ }}.051  &{\rm{\ }}.023  &{\rm{\ }}.016  &{\rm{\ }}.026  &{\rm{\ }}.015  &{\rm{\ }}.005  &{\rm{\ }}.003  &{\rm{\ }}.007  &{\rm{\ }}.004  \\
\hline
$\tau$ (anlys.) &{\rm{\ }}.053 &{\rm{\ }}.047 &{\rm{\ }}.055 &{\rm{\ }}.048 &{\rm{\ }}.025 &{\rm{\ }}.017 &{\rm{\ }}.028 &{\rm{\ }}.018 &{\rm{\ }}.006 &{\rm{\ }}.004 &{\rm{\ }}.007 &{\rm{\ }}.004 \\
\hline
\end{tabular}
\end{table}

After the RTS/CTS handshaking, more nodes are silenced,
refraining the nearby nodes from transmitting simultaneously.
The realistic backoff scheme along with the carrier sensing
should be considered to model ${{{\cal T}}_i}$ in \eqref{E:sir} properly.
By using MHP \cite{Stoyan1995},
the active transmitter density $\lambda_t$ in the contention-free period
can be modeled as follows:
\begin{equation}\label{E:density_rtscts}
{\lambda _t} = \frac{{1 - \exp \left( { - \lambda \tau \pi R_s^2} \right)}}{{\pi R_s^2}},
\end{equation}
where $R_s$ denotes the sensing range.
The value $\tau$ is the probability that an arbitrary transmitter in the network
completes the BEB process (i.e., backoff counter reaches 0) and accesses the channel.
Therefore, $\lambda\tau$ accurately represents the node density of contending nodes.
In this regard, we used a thinned node density $\lambda\tau$
instead of $\lambda$ to model the effect of BEB in \eqref{E:density_rtscts}.

The sensing range $R_s$ is a function of the physical carrier sensing threshold $I_s$.
In \cite{Hwang2011csma}, the authors used
the mean value of the sensing range, which is given as follows:
\begin{equation}\label{E:range_threshold}
{R_s}\!=\!{D_5}\!\int_0^{{D_0}}\!{f\left( r \right)dr}\!+\!\sum\limits_{i = 1}^5 {{D_{5 - i}}\!\int_{{D_{i - 1}}}^{{D_i}}\!{f\left( r \right)dr} }\!+\!{D_0}\!\int_{{D_5}}^\infty\!{f\left( r \right)dr},
\end{equation}
where $f\left( r \right)\!=\!2\lambda\tau \pi r\exp \left( { - \lambda\tau \pi {r^2}} \right)$
and ${D_i}\!=\!\sqrt[\alpha ]{{\left( {i + 1} \right)P/{I_s}}}$ for $i\!=\!0, \ldots ,5$.
The value $D_i$ means the minimum distance from an arbitrary node to the interferers.
It is clear that $R_s$ is in inverse proportion to $I_s$.
However, the dynamics between them is affected by the node density.
In the sparse node density case,
it is most probable that there is only one interferer nearby the sensing node.
As node density grows, at most six strong interferers can exist at the same distance
(refer to \cite{Hwang2011csma} for detail).
In this regard,
$R_s$ can be approximated as $R_s\!\approx\!D_0$
and $R_s\!\approx\!D_5$
for sparse and dense cases, respectively.
In next section, we will find the optimal carrier sensing threshold $I_s^*$.

\vspace{-4.5mm}

\section{Optimal Carrier Sensing Threshold}

\begin{figure}[tb]
\begin{center}
\includegraphics[width=3.3in]{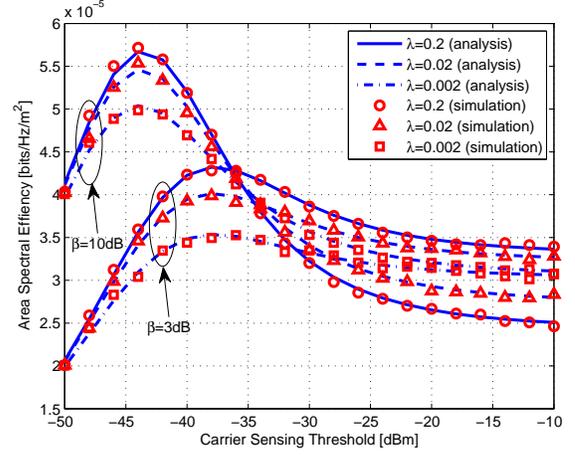}
\centering
\caption{Area spectral efficiency of CSMA/CA as a function of the
carrier sensing threshold ($W_0\!\!=\!\!16$, $m\!\!=\!\!32$ $r_t\!\!=\!\!50$~m, $P\!\!=\!\!30$~dBm).}
\label{F:ase_rtscts}
\end{center}
\end{figure}

When $I_s$ is high,
most transmitters are simultaneously transmitting,
making the success probability low. For lower $I_s$, more
transmitters are silent, and the aggregate interference is
less, leading to a higher success probability.
Thus, there exists an optimal carrier sensing threshold that
maximizes the ASE of \eqref{E:ase}.
Fig.~\ref{F:ase_rtscts} shows the ASE of the CSMA/CA
scheme as a function of $I_s$.
The simulation and our analytical results are congruent.
We observe that an optimal $I_s$
exists, which is obtained by solving \eqref{E:ase}.
To this end, the transmission success probability $p_s$
in \eqref{E:ase} is derived in the next subsection.

\vspace{-5mm}

\subsection{Transmission Success Probability of CSMA/CA}

With the carrier sensing range $R_s$, the other transmitters within
$R_s$ should be silenced.
The transmission of a typical transmitter is successful
if $\gamma_{j} \geq \beta$ is satisfied.
Assuming path loss exponent $\alpha$=4, which is validated for urban area,
the transmission success probability $p_s$ can be approximated in
closed-form as follows:
\begin{equation}\label{E:suc_pr_rtscts}
{p_s} \simeq \exp \left( { - \pi {\lambda _t}\sqrt \beta  r_t^2\arctan \left( {\frac{{\sqrt \beta  r_t^2}}{{R_s^2}}} \right)} \right).
\end{equation}
Details of the derivation are contained in Appendix.

\vspace{-5mm}

\subsection{Proposed Algorithm}

We now explain our main result for the optimal carrier sensing threshold.
Using \eqref{E:suc_pr_rtscts},
the ASE of \eqref{E:ase} is as follows:
\begin{equation}\label{E:ase1}
\begin{split}
\eta &= \lambda_t\log_2 \left( {1 + \beta } \right)
\exp \left( {-\pi \lambda_t\sqrt \beta  r_t^2\arctan \left( {\frac{{\sqrt \beta  r_t^2}}{{R_s^2}}} \right)} \right).
\end{split}
\end{equation}
Equation~\eqref{E:ase1} is a function of $R_s$, and
$R_s$ is a function of $I_s$ as shown in \eqref{E:range_threshold}.
 The value $\lambda_t$ is obtained using \eqref{E:density_rtscts}.
Unfortunately, the closed-form solution of \eqref{E:ase} is hard to
find. One way to deal with the problem is making an algorithm where the transmitters
update their sensing thresholds iteratively and distributively.
Our proposed algorithm is given as follows:
\begin{enumerate}
    \item First, initialize the carrier sensing threshold $I_s^{(0)}$
with a small value less than $r_t^{-\alpha}P$.
    \item Next, find the value $\tau$ using Newton's method \eqref{E:newton_tau}.
    \item Update $I_s^{(n)}$ using the following Newton's method:
\begin{equation}\label{E:ase_iterative}
I_s^{\left( {n + 1} \right)} = I_s^{\left( n \right)} - \frac{\eta'\left( {{I_s^{\left( n \right)}}} \right)}{\eta''\left( {{I_s^{\left( n \right)}}} \right)},
\end{equation}
where $\eta '$ and $\eta ''$ denote the first and second derivatives of
$\eta$ of \eqref{E:ase1} with respect to $I_s$.
    \item Repeat procedures 2) and 3) until the solution is found.
\end{enumerate}
The pseudo code of proposed algorithm is described in Algorithm~\ref{A:iter}.
The proposed algorithm converges to an optimal value
within a few iterations.
The transmission distance $r_t$ can be estimated by the received signal strength (RSS)
and/or Global Positioning System (GPS) information.
We conducted simulation based on the RSS method \cite{Chitte2009}.
The RSS measurements are relatively inexpensive and simple to implement in hardware.
To estimate the node density,
a node collects the received power samples from its nearest neighbors
and performs the maximum likelihood estimation.
According to \cite{Onur2012}, the estimation results are highly accurate
and the procedure is uncomplicated.

\begin{figure}[tb]
\centering
\subfloat[Optimal carrier sensing threshold]
     {\includegraphics[width=3.3in]{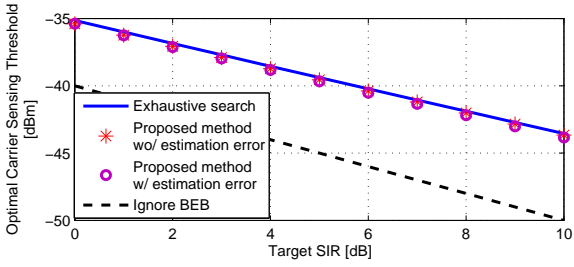}\label{F:optimal_sensing_threshold}}
\hfill  % vertical alignment
\centering
\subfloat[Maximum area spectral efficiency]
     {\includegraphics[width=3.3in]{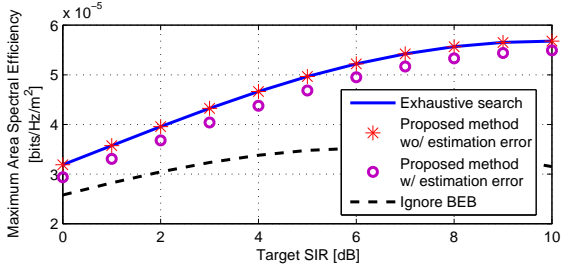}\label{F:optimal_ase}}
\centering
\caption{Optimal carrier sensing threshold and maximum area spectral efficiency
as a function of the target SIR
($\lambda\!\!=\!\!0.2$, $W_0\!\!=\!\!16$, $m\!\!=\!\!32$ $r_t\!\!=\!\!50$~m, $P\!\!=\!\!30$~dBm).}\label{F:optimal_thr_ase}
\end{figure}

As shown in Fig.~\ref{F:optimal_sensing_threshold}, the results of the iterative solution
and exhaustive search are coherent.
The optimal carrier sensing threshold varies with the target SIR $\beta$.
If $\beta$ increases, $I_s$ should be decreased
to lower the active transmitter density.
For the comparison, we also plot the optimal carrier sensing threshold
obtained by ignoring the BEB
(dashed line in Fig~\ref{F:optimal_thr_ase}).
In this case, the optimal carrier sensing range can be approximated as
$1.1278\sqrt[4]{\beta }r_t$ (derivation in Appendix).
By ignoring the BEB,
the active transmitter density is overestimated, where the
corresponding optimal carrier sensing threshold is
lower, causing performance degradation as shown in
Fig.~\ref{F:optimal_ase}. Fig.~2 shows the impact of estimation errors on $r_t$ and $\lambda$.
The $r_t\!=\!50$m is estimated by 54.8547m and the $\lambda\!=\!0.2$ is
estimated by 0.1911.
Even though we adopted rather primitive estimation methods,
the proposed algorithm shows acceptable performance.

\vspace{-2.5mm}

\section{Concluding Remarks}

We proposed a tractable approach for the optimal carrier sensing threshold
of the random CSMA/CA networks.
Most previous works using stochastic geometry overlooked the effect of the random backoff.
We considered the effect of the practical backoff scheme
and verified accuracy of our analysis by NS-3 simulations.
Our analytical results could be employed in
the design and optimization of high performing CSMA/CA networks.
The spectrum sensing based cognitive radio network (CRN)
is one of the viable applications of our work.
If the spectrum is sensed as available, the multiple secondary transmitters would
access concurrently, causing collisions.
To avoid this situation,
the CSMA/CA-based MAC protocol for the CRN is desirable.
Our results can be used to find an optimal spectrum sensing level.

\vspace{-2.5mm}

\section*{Appendix}

\subsubsection{Derivation of \eqref{E:suc_pr_rtscts}}

We denote $I_{R_s}$ as the aggregate interference from the outside of region
with the radius $R_s$.
Using the fact that the channel gain $g_{i,j}$
is an exponential random variable and
taking expectation of $I_{R_s}$, then $p_s$ is:
\begin{equation}\label{E:suc_pr1}
{p_s} \simeq \Pr \left[ {\frac{{{g_{i,j}} r_t^{ - \alpha }P}}{{{I_{{R_s}}}}} \geq \beta } \right] = {\mathbb{E}_{{I_{{R_s}}}}}\left[ {\exp \left( { - \frac{{ \beta r_t^\alpha }}{P}{I_{{R_s}}}} \right)} \right].
\end{equation}
By substituting $s\!=\!{ \beta r_t^\alpha }/{P}$, \eqref{E:suc_pr1} becomes
the Laplace transform of shot-noise process $I_{R_s}$.
Using the result of \cite{Baccelli2009}, \eqref{E:suc_pr1} is %expressed as:
\begin{align}\label{E:suc_pr3_}
  p_s \simeq \exp \left( { - 2\pi {\lambda _t}\int_{{R_s}}^\infty  {\left( {1 - {\mathbb{E}_G}\left[ {{e^{ - sGP{v^{ - \alpha }}}}} \right]} \right)vdv} } \right),
\end{align}
where $v$ is a dummy variable representing the distance to a random interferer.
Using the moment generating function of the exponential random variable, the probability $p_s$ is
\begin{align}\label{E:suc_pr3}
  p_s \simeq \exp \left( { - 2\pi {\lambda _t}\int_{{R_s}}^\infty  {\left( {\frac{\beta }{{\beta  + v^\alpha/r_t^\alpha}}} \right)vdv} } \right).
\end{align}
Assuming $\alpha$=4, closed-form is obtained as follows:
\begin{equation}\label{E:suc_pr4}
{p_s} \simeq \exp \left( { - \pi {\lambda _t}\sqrt \beta  r_t^2\arctan \left( {\frac{{\sqrt \beta  r_t^2}}{{R_s^2}}} \right)} \right).
\end{equation}

\subsubsection{Derivation of Optimal Sensing Threshold ignoring BEB}

Assuming high node density and $\alpha\!=\!4$,
the value $\lambda_t$ approximates $\lambda_t \approx 1/{\pi R_s^2}$.
The objective function of \eqref{E:ase} becomes:
\begin{equation}\label{E:ase_high_density}
\eta  = \frac{{\log_2 \left( {1 + \beta } \right)}}{{\pi R_s^2}}\exp \left( { - \frac{{\sqrt \beta  r_t^2}}{{R_s^2}}\arctan \left( {\frac{{\sqrt \beta  r_t^2}}{{R_s^2}}} \right)} \right).
\end{equation}
Differentiating \eqref{E:ase_high_density} with $R_s$
and simplifying exponential term:
\begin{equation}\label{E:diff_ase}
\frac{{\partial \eta }}{{\partial {R_s}}} \approx \frac{{2\log_2 \left( {1 + \beta } \right)}}{{\pi R_s^3}}\exp \left( { - \frac{{\beta r_t^4}}{{R_s^4}}} \right)\left( {\frac{{\beta r_t^4}}{{R_s^4 + \beta r_t^4}} + \frac{{\beta r_t^4}}{{R_s^4}} - 1} \right)=0.
\end{equation}
By solving \eqref{E:diff_ase}, $R_s^*$ is obtained:
\begin{equation}\label{opt_r_s}
R_s^* = {\left( {{0.5\left(1 + \sqrt 5\right) }\beta r_t^4} \right)^{1/4}} \approx 1.1278{\beta ^{\frac{1}{4}}}{r_t}.
\end{equation}

\end{document}